\documentclass[11pt]{article}
\usepackage{graphicx}
\usepackage{multirow}
\newcommand{\BABARPubYear}    {06}

\newcommand{\BABARConfNumber} {002}
\newcommand{\SLACPubNumber} {11967}

\input pubboard/babarsym
\newcommand{\bi}{\begin{itemize}}
\newcommand{\ei}{\end{itemize}}

\newcommand{\Bztoksphig}{\ensuremath{\Bzb\to\phi\KS\gamma}}
\newcommand{\Bztokzphig}{\ensuremath{\Bzb\to \phi\Kzb\gamma}}
\newcommand{\Bptokphig}{\ensuremath{\Bm\to \phi\Km\gamma}}
\newcommand{\Bmtokphig}{\ensuremath{\Bm\to \phi\Km\gamma}}
\newcommand{\Btokphig}{\ensuremath{\B\to \phi K\gamma}}

\newcommand{\mrec}{\ensuremath{m_{\rm rec}}}

\newcommand{\mmiss}{\ensuremath{m_{\rm miss}}}
\long\def\inst#1{\par\nobreak\kern 4pt\nobreak
    {\it #1}\par\vskip 10pt plus 3pt minus 3pt}

\begin{document}

%\linenumbers

{\pagestyle{empty}

%% \begin{flushleft}
%% \rm \babar$\;$Analysis Document \#1518, Version 7 \\
%% \end{flushleft}

\begin{flushright}
\babar-CONF-\BABARPubYear/\BABARConfNumber \\
%\babar-PUB-\BABARPubYear/\BABARPubNumber \\
SLAC-PUB-\SLACPubNumber \\
%%hep-ex/\LANLNumber \\
July 2006 \\
\end{flushright}

\par\vskip 5cm

% Title of the paper
\begin{center}
\Large \bf \boldmath
Measurement of $B$ Decays to $\phi K \gamma$
\end{center}
\bigskip

\begin{center}
\large The \babar\ Collaboration\\
\mbox{ }\\
\today
\end{center}
\bigskip \bigskip

% Abstract
\begin{center}
\large \bf Abstract
\end{center}
We measure the branching fraction of the radiative $B^{-}$ decay
${\cal B}(B^-\to \phi K^- \gamma)~=~(3.46 \pm 0.57
^{+0.39}_{-0.37})\times 10^{-6}$, and set an upper limit on the
radiative $\kern 0.18em\overline{\kern -0.18em B}{^0}$ decay ${\cal
B}( \kern 0.18em\overline{\kern -0.18em B}{^0} \to \phi \kern
0.2em\overline{\kern -0.2em K}{^0} \gamma )~<~2.71\times~10^{-6}$ at
90\% confidence level. We also measure the direct $C\!P$ asymmetry of
the $B^-\to \phi K^- \gamma$ \ mode $\calA_{C\!P} = (-26.4 \pm 14.3
\pm 4.8)\%$. The uncertainties are statistical and systematic,
respectively. These measurements are based on 207 $\mbox{\,fb}^{-1}$
of data collected at the $\Y4S$ resonance with the \mbox{\slshape
B\kern-0.1em{\smaller A}\kern-0.1em B\kern-0.1em{\smaller A\kern-0.2em
R}}\ detector.

\vfill
%\centerline{To be submitted to \jprBase D-RC.}
\begin{center}
Submitted to the 33$^{\rm rd}$ International Conference on High-Energy Physics, ICHEP 06,\\
26 July---2 August 2006, Moscow, Russia
\end{center}

\vspace{1.0cm}
\begin{center}
{\em Stanford Linear Accelerator Center, Stanford University,
Stanford, CA 94309} \\ \vspace{0.1cm}\hrule\vspace{0.1cm}
Work supported in part by Department of Energy contract DE-AC03-76SF00515.
\end{center}

\newpage

}% End of pagestyle{empty}

% Input author list file
%
\begin{center}
\small

The \babar\ Collaboration,
\bigskip

%% author list as of 01-Jul-2006 (596 authors)
%
{B.~Aubert,}
{R.~Barate,}
{M.~Bona,}
{D.~Boutigny,}
{F.~Couderc,}
{Y.~Karyotakis,}
{J.~P.~Lees,}
{V.~Poireau,}
{V.~Tisserand,}
{A.~Zghiche}
\inst{Laboratoire de Physique des Particules, IN2P3/CNRS et Universit\'e de Savoie,
 F-74941 Annecy-Le-Vieux, France }
% ...

\end{center}\newpage

% reset footnote counter
\setcounter{footnote}{0}

% The body of the paper starts here

%Introduction
   %Motivation

\begin{figure}
  \centerline{\includegraphics[width=0.5\textwidth]{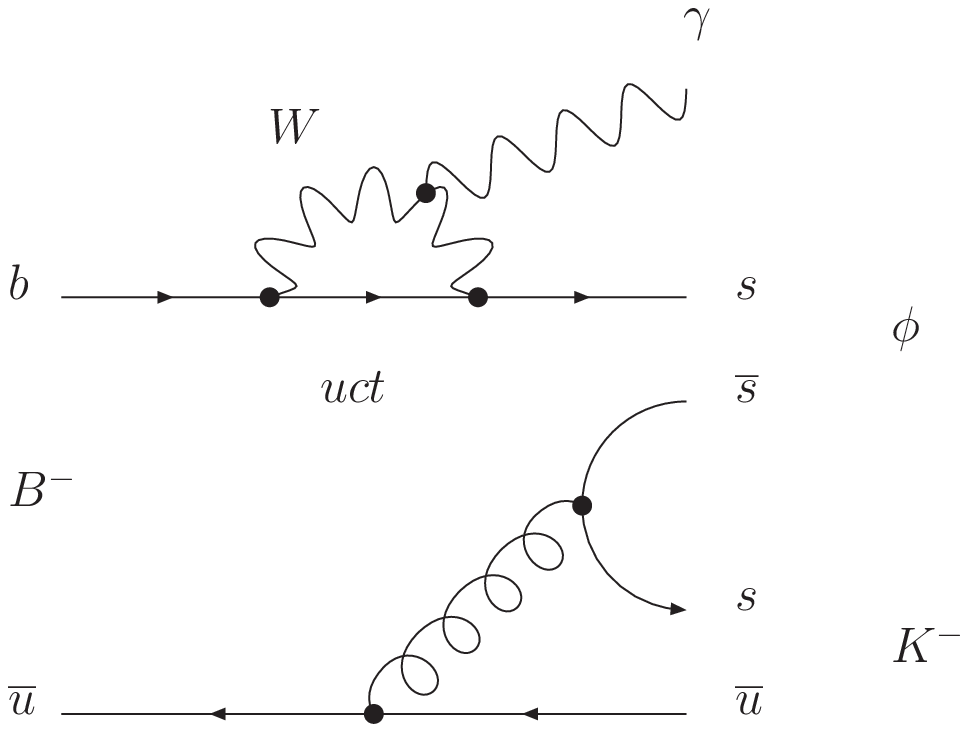}}
 \caption{A leading order penguin diagram for \Bmtokphig.
\label{fig:bsg-diagram}} 
\end{figure}

Measurements of the branching fractions and \CP asymmetries of
\btosgam decays provide a sensitive probe of the Standard Model (SM).
In the SM these decays are forbidden at tree level but allowed through
electroweak penguin processes (Fig.~\ref{fig:bsg-diagram}).  They are
therefore sensitive to the possible effects of new
physics~\cite{bsgteo} in the form of new heavy particles contributing
to the loop diagram. Additional contributions to the decay amplitudes
could affect branching fractions and \CP violation.  Furthermore, the
radiated photon is polarized due to the left-handed nature of the weak
interaction. The polarization can be probed by measuring the
time-dependent \CP asymmetry, which is sensitive to interference
between \Bz-\Bzb mixing and decay. Theoretical estimates in the
SM~\cite{bsg-update} bound the mixing-induced \CP asymmetry at about
the 10\% level. Here we focus on the time-integrated direct \CP
asymmetry, which is expected to be the same for charged and neutral
$B$ decays.

   %Context
Although exclusive \btosgam decays present a theoretical challenge
due to large non-perturbative QCD interactions, they are
experimentally clean. There have
already been results published for branching fraction and/or
\CP asymmetry measurements in several exclusive modes: $B \to \Kstar
\gamma$~\cite{kstar}, $\Bz \to \KS \piz \gamma$~\cite{kspi0g}, $B
\to \eta(') K \gamma$~\cite{ketagamma}, and various $B \to K \pi \pi
\gamma$~\cite{kpipi} modes. Here, we present a
measurement of the branching fraction for the charged mode \Bmtokphig\
and the neutral mode \Bztokzphig.~\footnote{Throughout this paper,
whenever a mode is given, the charge conjugate is also implied.} 
We also measure the direct \CP asymmetry in the
charged mode $\calA_{\CP} = [N(\Bm) - N(\Bp)]/[N(\Bm) + N(\Bp)]$, 
where the flavor of the $B$ is determined by the charge of the
kaon.
The Belle Collaboration has previously measured the branching fractions for
these modes, using $90 \invfb$ of \BB data at the \FourS
resonance~\cite{Drutskoy:2003}.
We describe the first \babar\ measurements of these modes using a
dataset that is more than twice as large.

%Describe any parts of the detector which are critical for the analysis.
The data used in this analysis were collected with the \babar\ detector
at the \pep2\ asymmetric \epem\ storage ring. This analysis is based on a
data set of $207 \invfb$ corresponding to 228 million \BB pairs collected
at the \FourS resonance. The \babar\ detector is described in detail
elsewhere~\cite{ref:babar}.
Important to this analysis are the tracking system composed of the
silicon vertex tracker (SVT) and drift chamber (DCH), the detector of
internally reflected Cherenkov light (DIRC), and
the electromagnetic calorimeter (EMC). The SVT and DCH
provide tracking and ionization energy loss (\dedx) measurements for
charged particles inside a
1.5 T magnetic field. The SVT is composed of five layers of double
sided silicon strips and covers a polar angle range between $20.1^o$
and $150.2^o$. The DCH continues tracking outside the SVT volume. It
consists of 40 layers of hexagonal cells filled with an 80:20 mixture
of helium:isobutane.  The DIRC is a ring imaging Cherenkov light
detector. Below 700~\mevc the tracking system provides most of the
charged particle identification (PID) information, while the DIRC
contributes more information at higher momenta.
Photons are detected and their energy measured in the
EMC, which is composed of 6580 thallium-doped CsI crystals.

% Event selection
We select events with EMC clusters of energy
$1.5-2.6~\gev$ in the \epem\ rest frame (CM frame)
that are not associated with any charged track. Photon
candidates are required to have an energy distribution consistent
with the shower shape of an electromagnetic interaction, and
they are required to be well isolated ($>~25~\cm$) from other calorimeter
clusters. A veto is applied to photon candidates that can be
combined with other neutral EMC clusters above a minimum threshold
energy to form an invariant mass consistent with a \piz
($115-155~\mevcc$) or $\eta$ ($470-620~\mevcc$). The threshold energy
is $50~\mev$ for \piz and $250~\mev$ for $\eta$.

We select $\phi$ candidates from pairs of oppositely charged kaon tracks,
determined not to be pions based on a PID
likelihood selection algorithm that uses \dedx and Cherenkov light
measurements. The same algorithm is used for the single \Kp in the
charged mode. The tracks are fitted to a vertex using a Kalman decay chain
fitter~\cite{Hulsbergen:2005pu}, and are required to
have a \chisq vertex probability greater than 0.1\%.
We select candidates with masses within a
$10~\mevcc$ window of the nominal $\phi$ mass~\cite{ref:pdg2004}.
In the neutral mode, pairs of oppositely charged tracks are fitted to
a common decay vertex and accepted as \KS candidates if the fit yields
a probability greater tha 0.1\%. The invariant mass of the pair is
required to be within $10~\mevcc$ of the \KS mass. The flight length
of is required to be greater than three times the uncertainty of that length.
We require the combined $\phi K$ invariant mass
to be less than $3.0~\gevcc$. In the neutral mode a
\Dz veto is applied by removing candidates with a $\phi K$ invariant mass
within $10~\mevcc$ of the \Dz mass.

The $K$, $\phi$, and $\gamma$ candidates are fitted to a common vertex
and accepted as $B$ candidates if the vertex
probability is greater than 0.1\%. To discriminate \BB events against
continuum background the ratio of Legendre moments
$L_2/L_0$ is required to be less than 0.55. The $L_i$ are defined by
$L_i = \sum_j |p^*_j| |\cos \theta^*_j |^i$,
where the $p^*_j$ are the CM momenta of all particles not
used in reconstructing the signal $B$ candidate, and the angle
$\theta^*_j$ is
between the particle's momentum and the thrust axis of the signal
$B$. We require the cosine of the angle between the $B$ candidate and the
beamline, $\cos \theta^*_B $, to be in the range [-0.9,0.9] in the CM frame.
We use two uncorrelated kinematic variables of the $B$ candidate:
the reconstructed mass \mrec\ and the missing mass
\mmiss. The reconstructed mass is the $B$ candidate invariant
mass calculated from the reconstructed energy and momentum. This is
required to be within $4.98-5.48~\gevcc$. The missing
mass squared is $\mmiss^2~=~\left(p_{\rm Beams}~-~p^{\rm mass\
const.}_B\right)^2$,  where $p_{\rm Beams}$ is the four-momentum of
the beams and $p^{\rm mass\ const.}_B$ is the four-momentum of the
$\Btokphig$ candidate after a mass constraint on the $B$ is applied.
We require the missing mass to be in the range $5.12-5.32~\gevcc$.

%Discuss Full Monte Carlo
To study event selection criteria we use simulated Monte Carlo (MC)
events of signal, generic $B$ decays, and $\epem \to \qqbar$ continuum
background, where $q = \{u,d,s,c\}$.
Events are generated using {\tt EVTGEN}~\cite{evtgen} and
the detector response simulated with {\tt GEANT4}~\cite{geant4}. Signal
Monte Carlo is generated according to the inclusive \btosgam scheme
presented in reference~\cite{Kagan99}, using $m_b = 4.62~\gevcc$ for
the effective $b$ quark mass. Exclusive signal MC is derived from this
by using only the part of the hadronic mass spectrum above the $\phi
K$ threshold of 1.52~\gevcc. Our selection criteria were chosen to
optimize the figure of merit $N_S/\sqrt{N_S + N_B}$ in the signal
region, where $N_S$ and $N_B$ are the number of signal and background
events, respectively, and the signal region is defined by $5.05 <
m_{\rm Rec} < 5.4 \gevcc$, $5.27 < \mmiss < 5.29 \gevcc$,
$|\cos \theta^*_B | < 0.8$, and $L_2/L_0 < 0.48$.

After all selection criteria are applied the average candidate multiplicity in
events with at least one candidate is approximately 1.01 and 1.07 in
the neutral and charged modes respectively.  If
multiple $B$ candidates are found in an event, we select the best one
based on a \chisq formed from the value and uncertainty of the mass of
the $\phi$ candidate and, in the neutral mode, the \KS candidate.  The
remaining background comes from continuum combinatorics and from other
$B$ decay modes, which will be discussed later.

% Fit description
We use an extended maximum likelihood fit in four observables -- 
\mmiss, \mrec, $L_2/L_0$, and $\cos \theta^*_B $ -- to extract the signal and
combinatoric background yields. 
The likelihood $\mathcal{L}$ is defined in the
following way: 
\begin{equation}
\mathcal{L} = \frac{ e^{-(N_S + N_B)}}{N!}
\Pi^N_i \left[ (N_S \mathcal{P}^i_S + N_B \mathcal{P}^i_B)\right].
\end{equation}
$N_S$ and $N_B$ are the fitted number of signal and background
events. $N$ is the total number of events used in the
fit. $\mathcal{P}^i_S$ and $\mathcal{P}^i_B$ are products of the
signal and background probability density functions (PDFs) for each
event $i$.  In the charged mode, in order to fit the \CP asymmetries
of the signal and the background, the numbers of \Bp and \Bm events
are determined separately: $N_j =
\frac{1}{2}(1+f\calA_{CP})n_{j}$, where $j=S$ or $B$;
$f$ is the flavor, defined as $+1$ for \Bm and $-1$ for \Bp;
$n_j$ and $\calA_{CP}^j$ are the total yield and \CP asymmetry of
species $j$. In the neutral mode $N_j=n_j$.

The signal PDFs for \mmiss\ and \mrec\ are parametrized as
asymmetric, variable-width Gaussian functions:
\begin{equation}
f(x) = exp \left[ \frac{-x^2}{2 \sigma^2_{L,R} +
\alpha_{L,R} x^2} \right].
\end{equation}
The parameters $\sigma_{L,R}$ and $\alpha_{L,R}$
determine the core width and variation of the width on either side of
$x=0$. The \mmiss\ background
PDF is an ARGUS function~\cite{argus}, with the endpoint calculated on
an event-by-event basis from the beam energy. The \mrec\ background
PDF is modelled as a $2^{\rm nd}$ degree polynomial.
The $L_2/L_0$
distribution is modelled using a binned PDF with eight bins, because
there is no {\em a priori} model for this distribution. There are
seven parameters in the PDF due to the condition that the bins sum to
unity. The signal and background models both use this form. The
$\cos \theta^*_B $ distribution is modelled as a $2^{\rm nd}$ degree
polynomial in both signal and background; for true $B$
candidates it is expected to follow $1 - \cos^2 \theta^*_B $.

%Control sample.
We use a  high statistics $\Bz \to \Kstarz(\to \Kp \pim) \gamma$
control sample to determine our signal shape parameters. Once
determined, these signal
parameters are fixed for the fit to \Btokphig\ data, while all
background shape parameters are allowed to vary. We fit for the number
of signal and background events, and in the charged mode the signal
and background \CP asymmetry as well. The same fit is used with signal
MC to determine the efficiency of the previously described selection
criteria. Corrections to the efficiency are discussed below.

% Yield and efficiency corrections
We apply several corrections to the fitted signal yield and efficiency
before determining the branching fractions. Studies of simulated
events show that our main sources of peaking backgrounds are
nonresonant $B \to K \Kp \Km \gamma$ events, and $B \to \phi K \piz$,
$B \to \phi K \eta$, where the \piz or $\eta$ decay fakes a high
energy photon.  We estimate the amount of $B \to K \Kp \Km \gamma$
contamination by fitting for the yield in $\phi$ mass sideband regions
extending outside the signal region from $10~\mevcc$ to $30~\mevcc$ of
the nominal $\phi$ mass. By interpolating into the signal region, we
find and correct for $0.03 \pm 1.5$ and $5.4 \pm 4.2$ events for the
neutral and charged modes respectively.  These contributions are
subtracted from the event yield found in the fit. We also subtract the
expected amount of $B \to \phi \Kstar(\to K \piz)$ as determined by
\babar~\cite{phikst}: 0.27 neutral and 1.98 charged events. Because
there have been no branching fraction measurements of $B \to \phi K
\piz$, $B \to \phi K \eta$, we assume that the branching fraction of
these modes is no more than three times that of $B \to
\phi\Kstar$. Therefore, we assign an uncertainty of 0.51 neutral and
2.86 charged events due to nonresonant $B \to \phi K (\piz/\eta)$
background.  To correct for any fit bias, we generate 1000
pseudo-experiments using our maximum likelihood PDFs with separate
components for \BB and continuum, and embedding signal events from the
full simulation. The background components are generated using shape
parameters determined from the full MC simulation.  We correct for a
bias of $+4.07 \pm 0.45$ events in the charged mode, due to
correlations among the observables in signal MC events that are not
accounted for in the fit. In the neutral mode we find a bias of $-0.06
\pm 0.20$, and so we include 0.20 events in the systematic uncertainty
of the yield.  We correct for known efficiency differences between
data and Monte Carlo in charged track, single photon, and \KS
reconstruction. These corrections amount to 0.956 in the neutral mode
and 0.975 in the charged mode. The corrected efficiencies are $(15.3
\pm 0.81)\%$ in the neutral mode and $(21.9 \pm 1.6)\%$ in the charged
mode, where the uncertainties are systematic (discussed below).

\begin{figure}
\begin{center}
\includegraphics[width=0.45\textwidth]{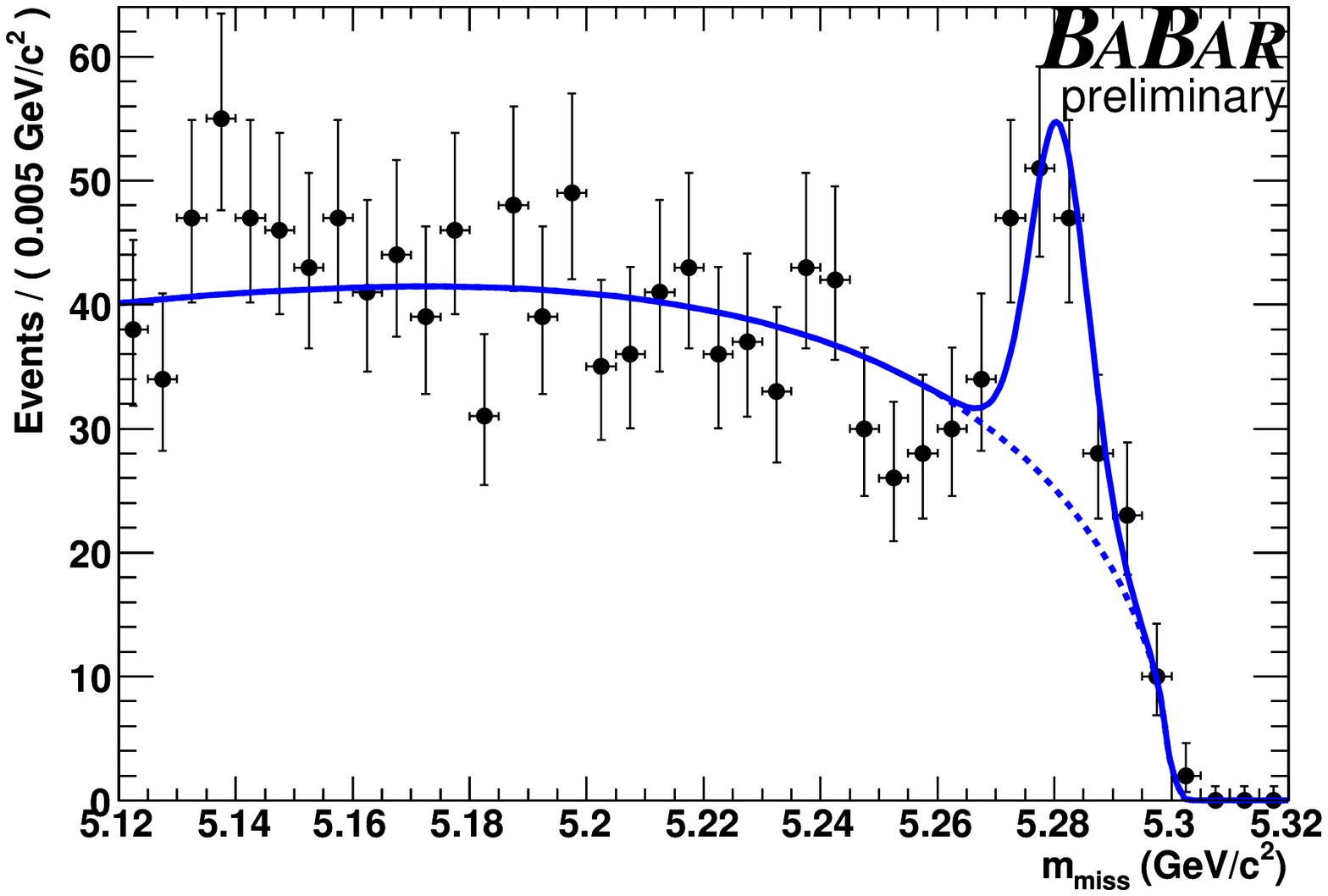}
\includegraphics[width=0.45\textwidth]{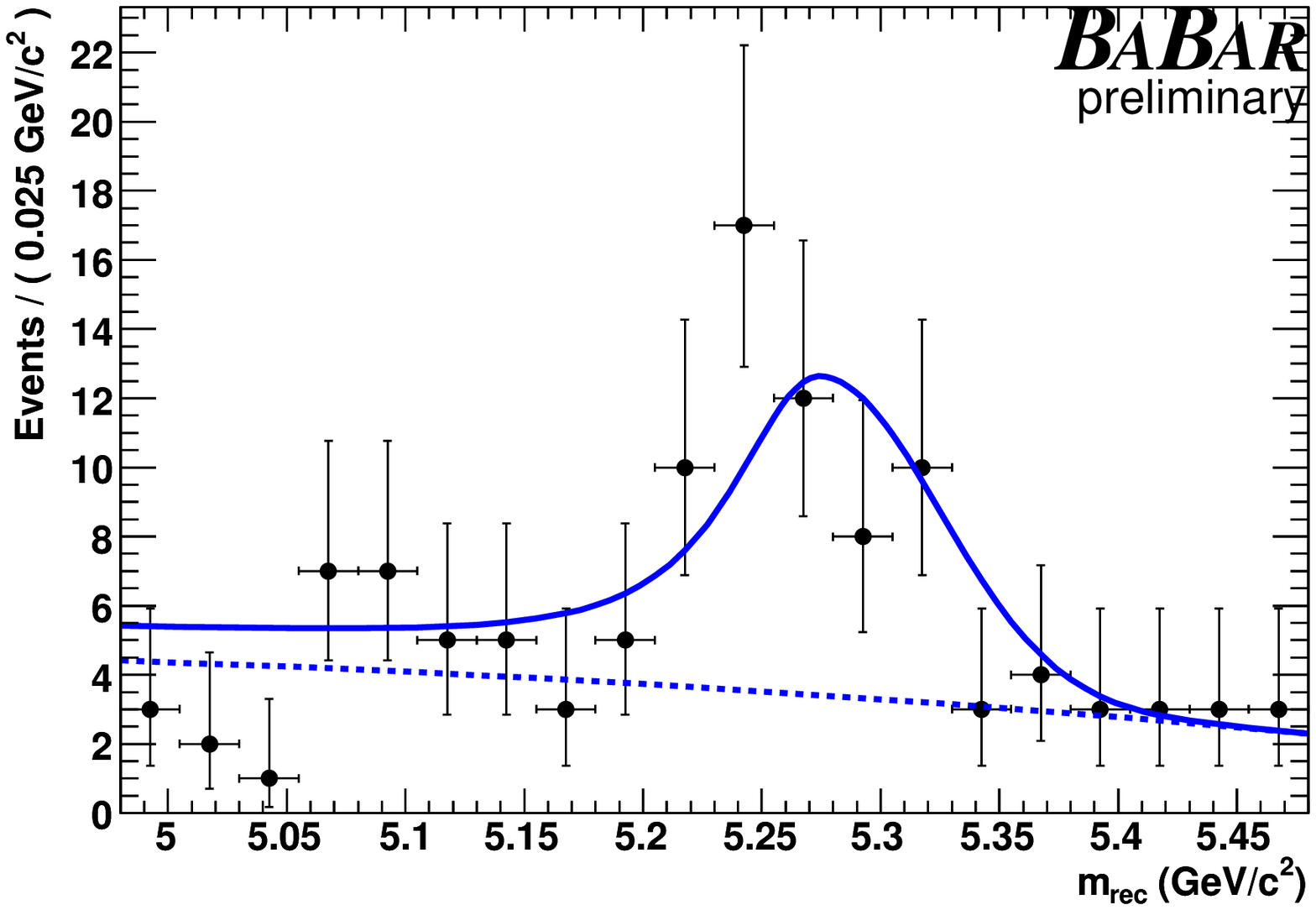}
\includegraphics[width=0.45\textwidth]{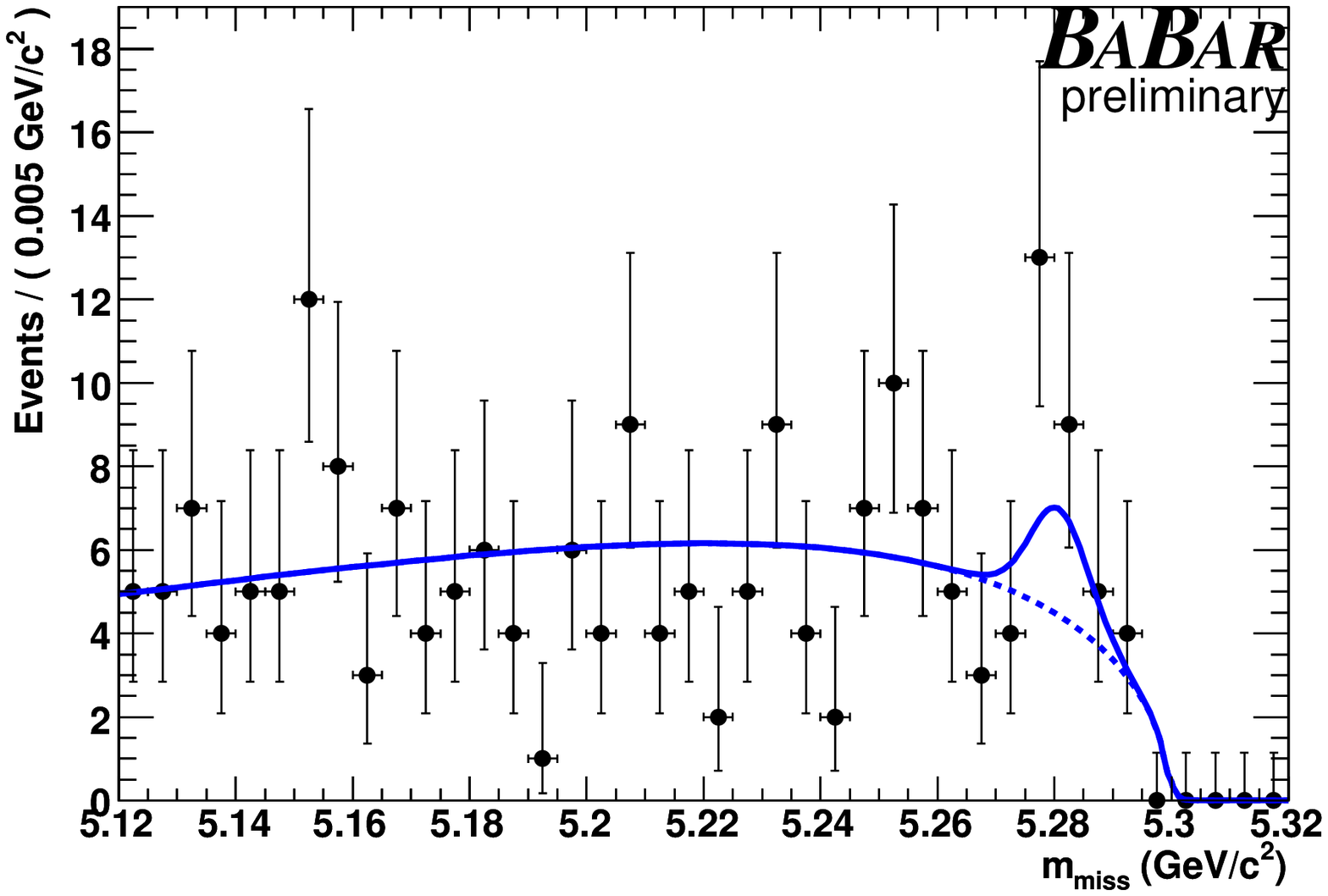}
\includegraphics[width=0.45\textwidth]{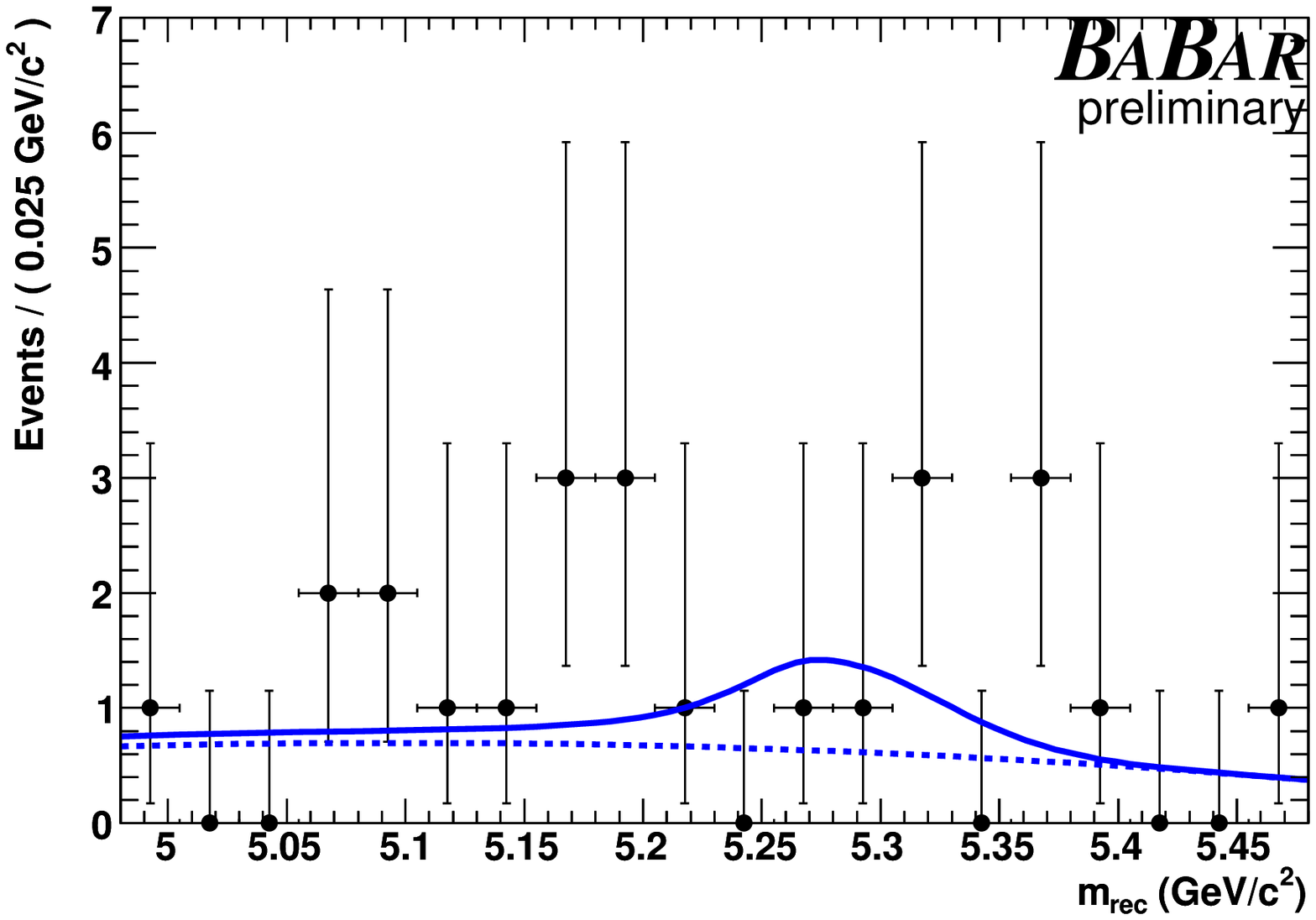}
\end{center}
\caption{Missing mass (left) and reconstructed mass (right) fit
projections in the signal region for the charged mode (upper) and the
neutral mode (bottom). The dotted curves show the fitted background
contribution while the solid curves show the signal.
\label{fig:fitresults} }
\end{figure}

% Results
The signal yields, efficiencies, branching fractions, and charged mode
\CP asymmetry are reported in Table~\ref{tab:final}. We calculate
the central value of the branching fractions as:
\begin{equation}
BF = \frac{N_{\rm sig}}{N_{\BB} \cdot \varepsilon \cdot \BR(\phi \ra
\Kp \Km) [\frac{1}{2} \BR(\KS \ra \pip \pim)]},
\end{equation}
where $N_{\rm sig}$ is the corrected number of signal candidates,
$N_{\BB} = (228.3 \pm 2.5) \times 10^6$ is the number of \BB pairs
recorded by \babar, $\varepsilon$ is the corrected efficiency, and
the (\KS \to \pip \pim) term is only used in the neutral mode. The partial
branching fractions are given by Ref.~\cite{ref:pdg2004}.
We measure $\BR(\Bztokzphig) = (1.35 \pm 0.92 ^{+0.31}_{-0.30}) \times
10^{-6}$ and $\BR(\Bmtokphig) = (3.46 \pm 0.57 ^{+0.39}_{-0.37})
\times 10^{-6}$. In the charged mode we measure
$\calA_{\CP} = (-26.4 \pm 14.3 \pm 4.8)\%$. Fits
to the missing mass and reconstructed mass distributions, projected
into the signal region defined earlier, are shown in
Figure~\ref{fig:fitresults}. We use a set of 1000 pseudo-experiments
in the neutral mode to determine the probability of
obtaining a branching fraction less than or equal to our measured
central value under the hypothesis that it is in fact the same as the
charged mode. This was found to be 1.1\%. 

For the neutral mode we compute the 90\% confidence level upper limit
on the branching fraction. We use a Bayesian approach with an {\em a
priori} probability for the branching ratio which is flat in the
physical region $0 \leq \BR \leq 1$, and zero elsewhere.
% Steve says unnecessary:
%The upper bound of the interval $\BR^{UB}$ is defined in
%$ \int_{0}^{\BR^{UB}}\mathcal{L(\BR)}dN /
%\int_{0}^{\inf}\mathcal{L(\BR)}d\BR  = 90\%,$
%where $\mathcal{L}$ is the likelihood of the measured sample of events.
The value of the likelihood function is computed by fixing the signal
yield to a desired value and fitting the other free parameters on
the data sample. The function is then integrated numerically.
We account for systematic uncertainties on the yield by  convolving
the likelihood function with the distribution of the
errors, before computing the upper limit.
To determine the upper limit of the signal yield we use a Gaussian PDF
having a width equal to the
systematic uncertainty of the yield. Similarly for the efficiency
uncertainty we use a Gaussian PDF having a width equal to the
systematic error. After also applying the yield corrections discussed
previously we obtain $\BR(\Bztokzphig) < 2.71 \times 10^{-6}$.

\begin{table}
  \centerline{
    \begin{tabular}{c|cccc}
      \hline\hline
      Channel      & Yield & Efficiency & $\BR (10^{-6})$ & $\calA_{CP}$\\
      \hline
\Bptokphig  & $85.0 \pm 13.9 ^{+7.3}_{-6.9}$&$[21.9\pm1.6\syst]\%$ & $3.46 \pm 0.57 ^{+0.39}_{-0.37}$ & $(-26.4 \pm 14.3 \pm 4.8)$\% \\
\multirow{2}{*}{\Bztokzphig} & $8.0 \pm 5.5 ^{+1.8}_{-1.7}$ & \multirow{2}{*}{$[15.33\pm0.81\syst]\%$} & $1.35 \pm 0.92 ^{+0.31}_{-0.30}$ & \\
            & $<16.0$           &                       & $<2.71$ & \\
      \hline \hline
    \end{tabular}
  }
  \caption{ Summary of the branching fractions and direct \CP
asymmetry. In \Bztokzphig\ the 90\% confidence level upper limit is
also given.
    \label{tab:final} }
\end{table}

\begin{figure}
\begin{center}
\includegraphics[width=0.45\textwidth]{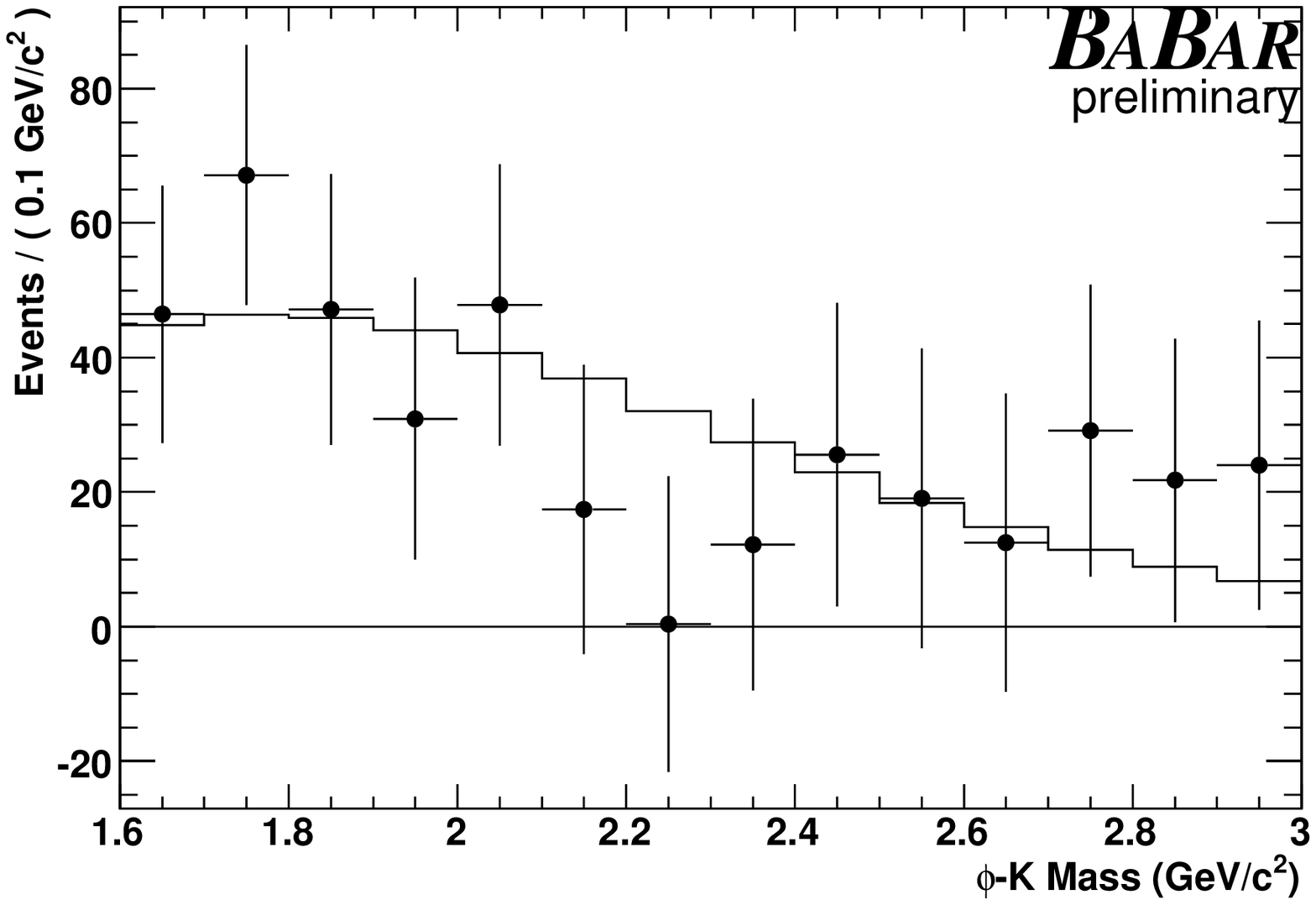}
\includegraphics[width=0.45\textwidth]{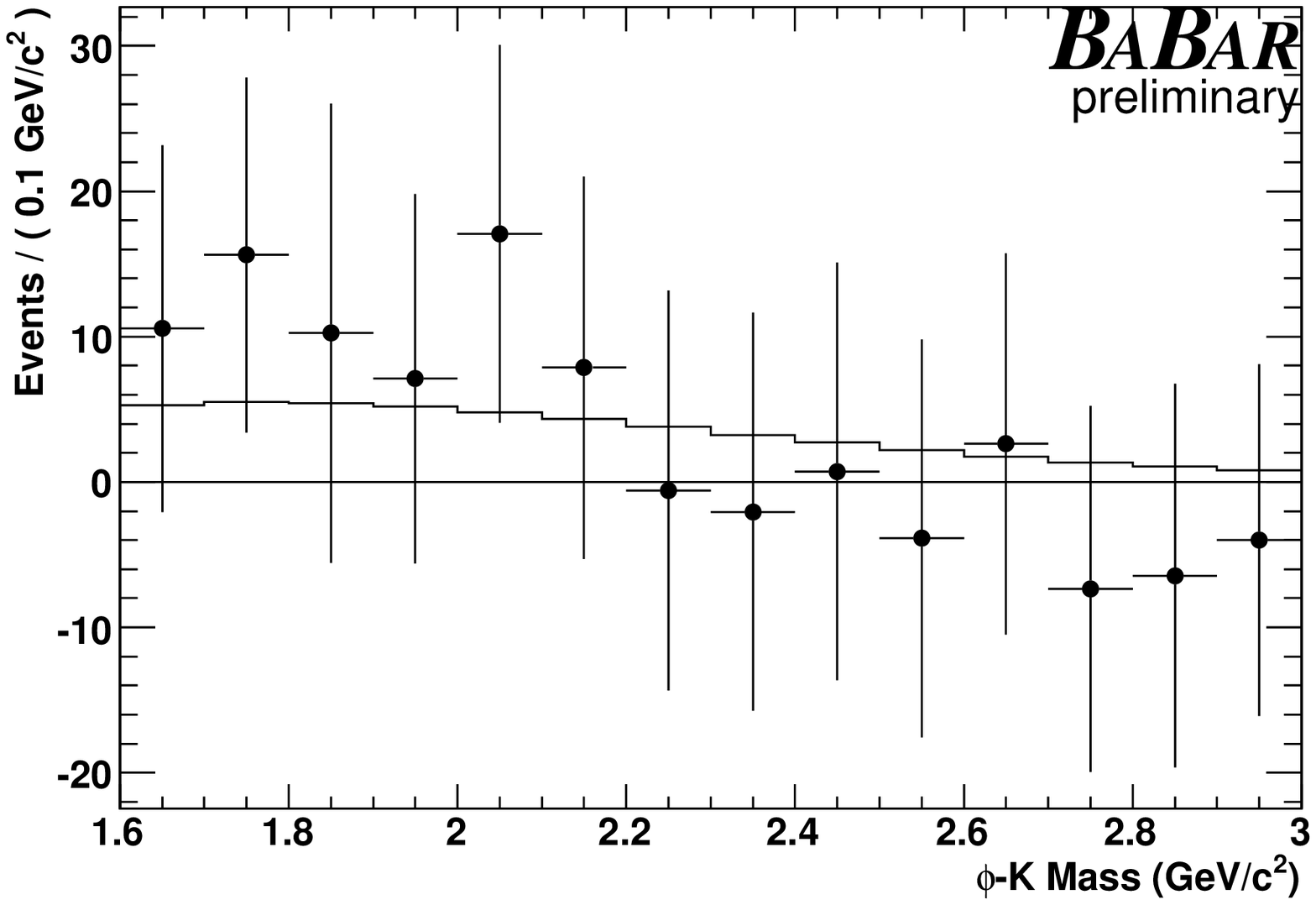}
\end{center}
\caption{The background-subtracted and efficiency-corrected $\phi K$
mass distributions
(points with uncertainties) for the charged mode (left) and the
neutral mode (right). The signal MC prediction for the mass spectrum, based on
Ref. ~\cite{Kagan99}, is shown as a histogram without uncertainties.
\label{fig:phikmass} }
\end{figure}

To determine the contribution from resonances decaying to $\phi K$ we
study the background-subtracted~\cite{splot}, efficiency corrected
$\phi K$ invariant mass
distribution, shown in Figure~\ref{fig:phikmass}. Using the charged mode, we
find that no more than 50\% of the spectrum in the $1.6-3.0~\gevcc$
range comes from the $K_2(1770)$ resonance. We use this to bound our
model uncertainty, described below.

% Cross checks/systematics
We assign an uncertainty due to the fixed signal parameters in
the fit. The parameters were obtained from the control
sample, and therefore they have some statistical uncertainty. We
varied these parameters within their uncertainties to determine the
total uncertainty on the yields.
We account for other
systematic uncertainties in charged kaon tracking, kaon PID, \KS,
$\phi$, and photon selection efficiency. There are small
uncertainties assigned with our $L_2/L_0$ selection and the \piz /
$\eta$ veto. To account for uncertainty due to the assumption of a
specific $\phi K$ mass spectrum
for simulated events, we determine what our efficiency would have been
in the scenario that half of the spectrum comes from $K_2(1770)$
resonant $\phi K$ production, while the other half comes from our
signal MC model. We assign the relative efficiency difference as an
uncertainty. Adding all of the uncertainties in quadrature, we
find a total acceptance/efficiency uncertainty of 5.2\% in the neutral mode and
7.1\% in the charged mode. The contributions for each mode are
summarized in Table~\ref{tab:syst-final}.

\begin{table}
  \centerline{
    \begin{tabular}{c|c|c}
      \hline\hline
             & \Bztoksphig     & \Bmtokphig \\
      Source & Uncertainty & Uncertainty \\
      \hline
      $K\ \Kp \Km \gamma$ subtraction  & 19.7\% & 5.2\% \\
      Peaking Background           & 6.4 \% & 3.4\%  \\
      Fit Bias                     & 2.6 \% & 0.6\%  \\
      Fit PDF parameters           & $^{+7.0}_{-5.9}$ \% & $^{+5.9}_{-5.2}$ \%  \\
      \hline
  \bf  Total yield uncertainty   & $\bf ^{+1.8}_{-1.7}$ events & $\bf ^{+7.3}_{-6.9}$ events \\
      \hline
      Kaon Tracking                & 2.8\% & 4.2\%  \\
      \KS Efficiency               & 1.5\% & 0\%    \\
      $\phi$ Efficiency            & 1.7\% & 1.7\%  \\
      Particle ID                  & 2.8\% & 4.2\%  \\
      Single Photon Efficiency     & 1.8\% & 1.8\%  \\
      Photon Spectrum Model        & 0.4\% & 2.6\%  \\
      $L_2/L_0$ Cut                & 1.2\% & 1.2\%  \\
      $\piz / \eta$ Veto           & 1\%   & 1\%    \\
      \hline
\bf   Efficiency/acceptance uncertainty  & $\bf 5.2\%$ & $\bf 7.1\%$ \\
      \hline 
\bf   \BB Counting                 & \bf 1.1\%   & \bf 1.1\% \\
      \hline
      \hline
\bf   Total                        & $\bf ^{+23}_{-22}$ \% & $\bf ^{+11.2}_{-10.8}$\%\\
      \hline \hline
    \end{tabular}}
     \caption{Summary of the systematic uncertainties.
	\label{tab:syst-final} } 
\end{table}

For the direct \CP asymmetry measurement we assume that the efficiency
corrections and uncertainties cancel out. To account for uncertainty 
due to peaking background we use the following procedure. We assume an
{\it a priori} flat distribution for the \CP asymmetry between $-1$ and 
1, which has a root mean square width of 0.58. We multiply this by the
expected fractional contamination in our sample to obtain the 
systematic uncertainty. For $\Bm \to \phi \Km
(\piz/\eta)$ we assign 1.8\% uncertainty, while for $\Bm \to \Km \Kp 
\Km \gamma$ we assign 3.5\% uncertainty. For resonant $B \to \phi \Kstar(\to K
\piz)$ events, the previous \babar\ measurement~\cite{phikst} shows
that the \CP asymmetry is consistent with zero to within 9\%. We
therefore consider this to be negligible in our case.  Using the
control sample as we did with the branching fraction measurement, we
vary the fixed input parameters of the fit to determine the
uncertainty on the signal \CP asymmetry. This was found to be 2.2\%,
bringing the total systematic uncertainty to 4.8\%.   

In summary, we have performed the first \babar\ studies of \Btokphig\ decay
modes. The \Bptokphig\ branching fraction was measured, and an upper
limit for the \Bztokzphig\ mode was determined. Our measurements are
consistent with the assumption of isospin symmetry at the 1.1\%
level. We have made the first measurement of the direct \CP asymmetry in
\Bmtokphig. For comparison, we quote the Belle
results~\cite{Drutskoy:2003}: $\BR(\Bmtokphig) = (3.4 \pm 0.9 \pm 0.4)
\times 10^{-6}$ and $\BR(\Bztokzphig) < 8.3 \times 10^{-6}$ at 90\%
confidence level.
The statistical uncertainties will improve as the $B$ Factories collect
more data over the next several years. In future measurements a large
amount of \CP violation would be a sign of physics beyond the Standard Model.

% Specific acknowledgments for this paper; remove if not needed.

% Standard acknowledgments paragraph; must always be included.
We are grateful for the 
extraordinary contributions of our \pep2\ colleagues in
achieving the excellent luminosity and machine conditions
that have made this work possible.
The success of this project also relies critically on the 
expertise and dedication of the computing organizations that 
support \babar.
The collaborating institutions wish to thank 
SLAC for its support and the kind hospitality extended to them. 
This work is supported by the
US Department of Energy
and National Science Foundation, the
Natural Sciences and Engineering Research Council (Canada),
Institute of High Energy Physics (China), the
Commissariat \`a l'Energie Atomique and
Institut National de Physique Nucl\'eaire et de Physique des Particules
(France), the
Bundesministerium f\"ur Bildung und Forschung and
Deutsche Forschungsgemeinschaft
(Germany), the
Istituto Nazionale di Fisica Nucleare (Italy),
the Foundation for Fundamental Research on Matter (The Netherlands),
the Research Council of Norway, the
Ministry of Science and Technology of the Russian Federation, 
Ministerio de Educaci\'on y Ciencia (Spain), and the
Particle Physics and Astronomy Research Council (United Kingdom). 
Individuals have received support from 
the Marie-Curie IEF program (European Union) and
the A. P. Sloan Foundation.

\end{document}